\documentclass[prl,twocolumn,showpacs,amsmath,amssymb,superscriptaddress]{revtex4-1}

\usepackage{graphicx}
\usepackage{bm}
\usepackage[usenames,dvipsnames,svgnames,table]{xcolor}
\usepackage[colorlinks=true,linkcolor=RoyalBlue,citecolor=RoyalBlue]{hyperref}
\usepackage{soul}

\begin{document}
\title{Stacking-Dependent Magnetism in Bilayer CrI$_3$}

\author{Nikhil Sivadas} 
\email{ns462@cornell.edu}
\affiliation{School of Applied and Engineering Physics, Cornell University, Ithaca, New York 14853, USA}

\author{Satoshi Okamoto} 
\affiliation{Materials Science and Technology Division, Oak Ridge National Laboratory, Oak Ridge, Tennessee 37831, USA}

\author{Xiaodong Xu} 
\affiliation{Department of Physics, University of Washington, Seattle,WA98195, USA}
\affiliation{Department of Materials Science and Engineering, University of Washington, Seattle, WA 98195, USA}

\author{Craig.~J.~Fennie} 
\affiliation{School of Applied and Engineering Physics, Cornell University, Ithaca, New York 14853, USA}

\author{Di Xiao}  
\affiliation{Department of Physics, Carnegie Mellon University, Pittsburgh, Pennsylvania 15213, USA}
\date{\today}

\begin{abstract}
We report the connection between the stacking order and magnetic properties of bilayer CrI$_3$ using first-principles calculations.  We show that the stacking order defines the magnetic ground state. By changing the interlayer stacking order one can tune the interlayer exchange interaction between antiferromagnetic and ferromagnetic. To measure the predicted stacking-dependent magnetism, we propose using linear magnetoelectric effect. Our results not only gives a possible explanation for the observed antiferromagnetism in bilayer CrI$_3$ but also have direct implications in heterostructures made of two-dimensional magnets.

$\bm{KEYWORDS:}$ 2D magnets, Moir\'{e} superlattices, stacking order, CrI$_3$, super-superexchange, beyond graphene
\end{abstract}
\maketitle

Since the demonstration of intrinsic ferromagnetism (FM) in atomically thin crystals~\cite{Huang17p270,Gong17p265}, there has been a lot of interest in two-dimensional (2D) magnets~\cite{Xing17p024009, Lee16p7433,Seyler18p277,PhysRevLett.120.136402,Lin18p072405,Shcherbakov18p4214,Bonilla18p41565}.
Among them, CrI$_3$ presents an intriguing case.  While bulk CrI$_3$ is FM, it becomes a layered antiferromagnet (AFM) when thinned down to a few atomic layers~\cite{Huang17p270}.  A number of interesting phenomena associated with this layered antiferromagnetism have been observed, including giant tunneling magnetoresistance when CrI$_3$ is used as the tunnel barrier~\cite{Song18p4851,Klein18p3617,Wang18p2516}, and gate tunable magneto-optical Kerr effect, along with electrostatic doping control of magnetism~\cite{Sivadas16p267203,Jiang18p41563,Huang18p544,Jiang18p41563,Jiang18p549}. However, despite the huge interest, the origin of the AFM interlayer exchange in bilayer CrI$_3$ remains unclear.

Motivated by the above question, in this Letter we explore the connection between the crystal structure and magnetic properties of bilayer CrI$_3$ using first-principles calculations.
We find that the stacking order defines the magnetic ground state.
The coupling of stacking order and magnetism is qualitatively unaffected by atomic relaxation, and is, therefore robust. This stacking-dependent magnetism originates from the competition between orbital-dependent interlayer AFM super-superexchange (SSE) and interlayer FM SSE.
Thus, by changing the stacking order one can tune the magnetic ground state between AFM and FM.  
We also propose using linear magnetoelectric (ME) effect to distinguish between the various predicted AFM stacking orders. In addition to providing a possible explanation for the observed AFM in bilayer CrI$_3$, our results have a broader impact on other 2D honeycomb magnets such as Cr$X_3$ ($X = $ Cl, Br, I)~\cite{Handy52p891,Bengel95p95,Sivadas15p235425,McGuire17p064001} and their heterostructures, including magnetic Moir\'{e} superlattices.

\begin{figure*}
\includegraphics[width=1.0\textwidth]{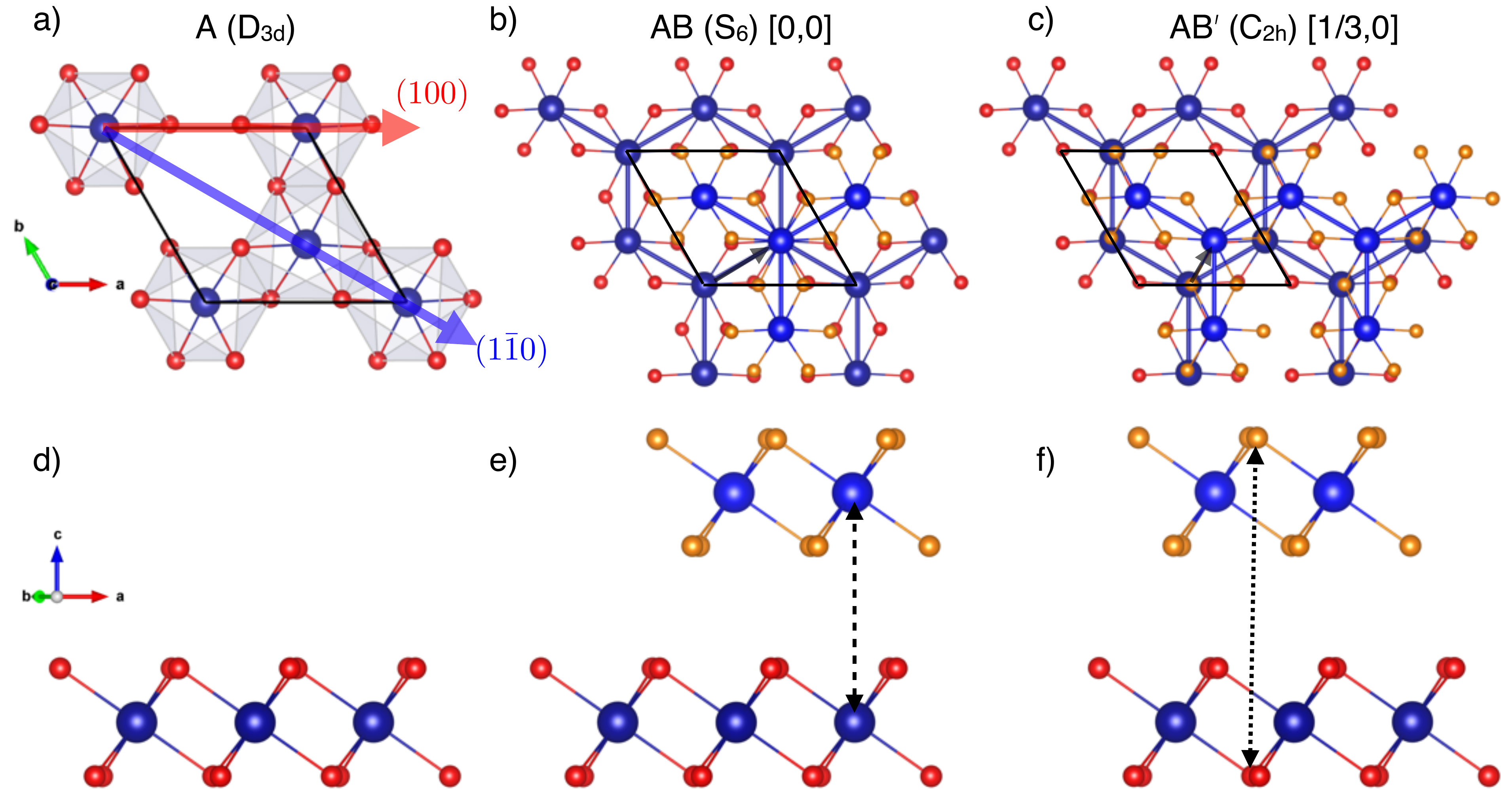}
\caption{\label{fig:Cry} The crystal structure for different stacking orders of CrI$_3$. (a)-(c) The top view, and (d)-(f) the side view of monolayer CrI$_3$ in (a) and (d), bilayer CrI$_3$ in AB-stacking in (b) and (e), and bilayer CrI$_3$ in AB$^{\prime}$-stacking in (c) and (f), along with their corresponding point groups. For clarity, the same atoms in the top and the bottom layer are shown by different colors. Blue (dark blue) balls represent the Cr atoms in the top (bottom) layer. Orange (red) balls represent the I atoms in the top (bottom) layer. The Cr-I octahedron is shown only for the monolayer. The high-symmetry [100]- and [1$\bar{1}$0]- lateral shift directions are labeled in (a) as red and blue translucent arrows, respectively. The honeycomb network of the Cr atoms is shown for the two bilayers in (b) and (c). The relative stacking between the layers is shown by black arrows. AB$^{\prime}$-stacking corresponds to a fractional lateral shift of the bottom CrI$_3$ layer by [1/3,0] with respect to the AB-stacking, and is labeled in (c). The change in the relative stacking order for the lateral shift is shown in (e) and (f) by dashed and dotted lines. The crystal structure is drawn using VESTA~\cite{Momma11p1272}.}
\end{figure*}

As our goal is to understand the stacking dependence of magnetic order in CrI$_3$, we begin our discussion with the crystal structure of CrI$_3$. Monolayer CrI$_3$ consists of magnetic Cr ions that form a honeycomb lattice, with each Cr atom coordinated by six I atoms that form a distorted edge-sharing octahedron [see Fig.~\ref{fig:Cry} (a), (d)]. Monolayer CrI$_3$ has the point group D$_{3d}$. Bulk CrI$_3$ can be obtained by stacking these monolayer units, which we label as the `A'-block [see Fig.~\ref{fig:Cry} (a)]. The bulk has a low-temperature (below 210~K) phase with the space group R$\bar{3}$ and a room-temperature phase with space group C2/m~\cite{Handy52p891, McGuire15p612, Jiang18p41563}. The low-temperature phase has an ABC-Bernal stacking with each layer laterally shifted by [2/3, 1/3] in fractional coordinates with respect to the neighboring layer. The reduction in symmetry for the high-temperature phase is associated with a relative lateral shift of the stacking order, leaving the monolayer units unaffected~\cite{McGuire15p612}. This corresponds to a further lateral shift of [1/3, 0] and [2/3, 0] for the B- and C-monolayer units, respectively, with respect to the ABC-stacking. Thus, both R$\bar{3}$ and C2/m phases have an ABC-stacking sequence. To distinguish the two cases, we refer to the R$\bar{3}$ phase as the ABC-stacking sequence, and the C2/m phase as the AB$^{\prime}$C$^{\prime}$-stacking sequence. While considering the stacking order in the bilayer, it is, therefore, necessary to consider at least two high-symmetry stacking orders shown in Fig.~\ref{fig:Cry}: the AB-stacking (S$_6$ point group) from the low-temperature bulk structure [see Fig.~\ref{fig:Cry} (b), (e)], and the AB$^\prime$-stacking (C$_{2h}$ point group) from the high-temperature bulk structure [see Fig.~\ref{fig:Cry} (c), (f)]. Hereafter, the various stacking orders will be discussed with respect to the AB-stacking. 

We calculate the total energies using first-principles calculations as implemented in Vienna \textit{ab~initio} simulation package (VASP)~\cite{Kresse96p11169}, with the PBEsol functional~\cite{Perdew08p136406}. A Hubbard on-site Coulomb parameter ($U$) of 3~eV was chosen for the Cr atoms to account for strong electronic correlations, as suggested by Liechtenstein {\it et al}~\cite{Liechtensteinp95467}. A vacuum region in excess of 15~{\AA} was used for the bilayer. Structural relaxation was done with a force convergence tolerance of 1 meV/\AA, in a regular 12 $\times$ 12 $\times$ 1 Monkhorst-Pack grid with a plane-wave cutoff energy of 450~eV. The computed structural parameters are close to the experimental values for bulk CrI$_3$ and are listed in the Supporting Information~\cite{sup-Sta-dep}. A stacking-constraint relaxation was performed along the high-symmetry lateral shift directions, relaxing the atomic positions as well as the unit cell volume. We have also verified that the results presented are qualitatively independent of the choice of functional~\footnote{We checked this using PBE-functional with the Tkatchenko-Scheffler correction scheme (DFT-TS) as implemented in VASP~\cite{Tkatchenko09p073005,Tomas13p064110}, and the results are unaffected.}. We make use of ISOTROPY software suite to aid with the group-theoretic analysis~\cite{isotrophy,Campbell06p607}.  

\begin{figure}
\includegraphics[width=1.0\columnwidth]{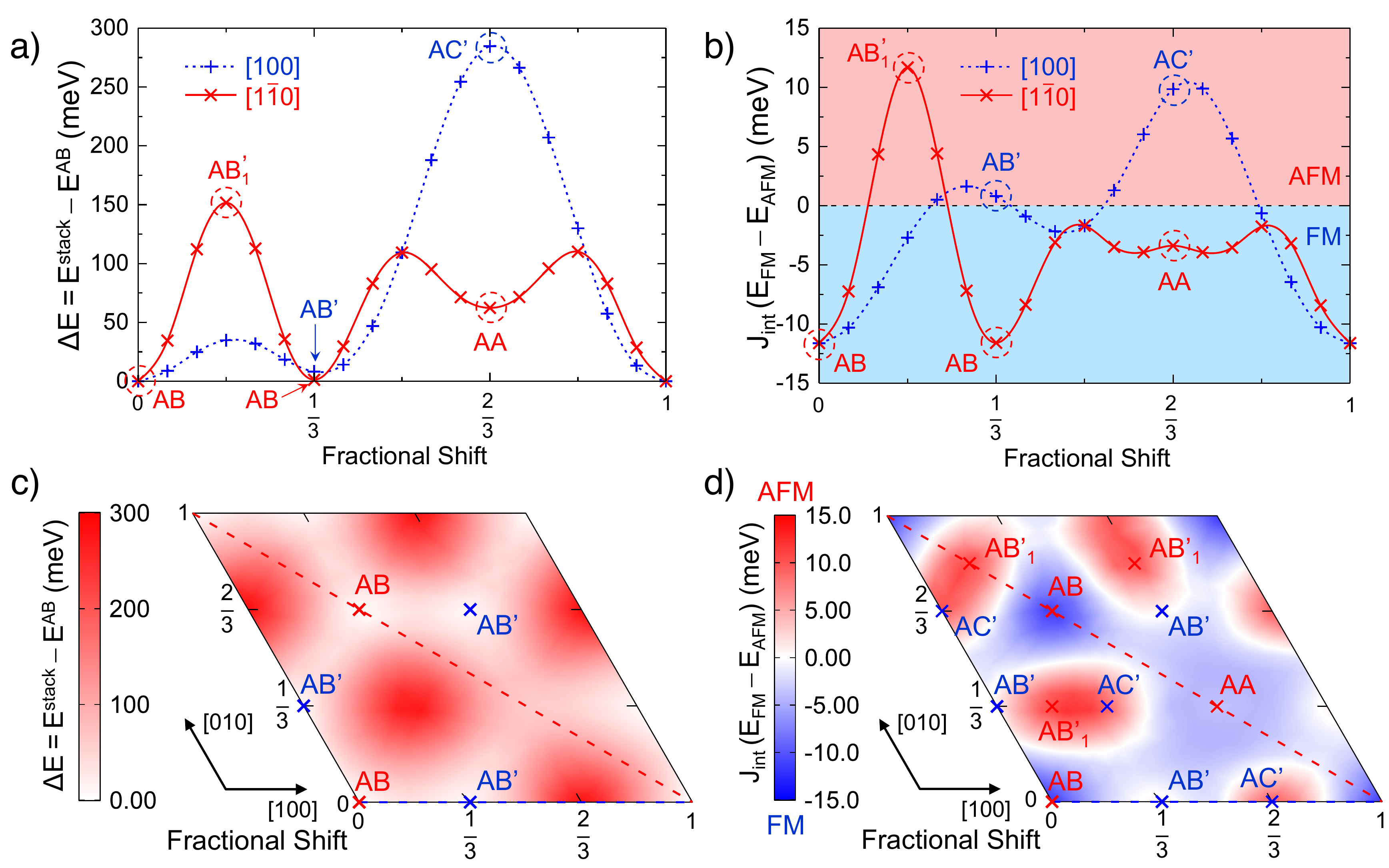}
\caption{\label{fig:exc} The stacking energy and the interlayer exchange energy as a function of lateral shift, with respect to the AB-stacking. The stacking energies are shown for (a) the high-symmetry [100]- (dotted blue line) and [1$\bar{1}$0]-directions (solid red line), and for (c) the full space of lateral shifts. The corresponding exchange energy is shown for different stacking orders for (b) the high-symmetry [100]- (dotted blue line) and [1$\bar{1}$0]-directions (solid red line), and for (d) the full space of lateral shifts, with positive (negative) regions corresponds to stacking which is AFM (FM).  AB$^\prime$-, AC$^\prime$-, AA- and AB$^\prime_1$-stacking orders correspond to a fractional lateral shift of the bottom CrI$_3$ layer by [1/3,0], [2/3,0], [2/3, 1/3] and [1/6, -1/6] with respect to the AB-stacking, respectively. The heat-maps in (c) and (d) were drawn by interpolating over the neighboring data points on a 6 $\times$ 6 grid.}
\end{figure}

Fig.~\ref{fig:exc}(a) shows the stacking energy defined as the change in energy of the ground state magnetic configuration for a particular stacking order, with respect to the AB-stacking, rigidly shifted along the high-symmetry [100]- (blue dotted line) and [1$\bar{1}$0]-directions (red solid line). A lateral shift along the [100]-direction corresponds to shifting a monolayer unit along the in-plane projection of the nearest Cr-I bond, whereas, a lateral shift along the [1$\bar{1}$0]-direction corresponds to shifting a monolayer unit along the nearest Cr-Cr bond. Naturally, these two directions are inequivalent. The shape of this stacking-dependent energy we find is similar to that of hexagonal boron nitride and graphene along the high-symmetry directions~\cite{Marom10p046801,Constantinescu13p036104,popov12p82}. 

We note that while AB-stacking is at the global minimum, AB$^\prime$-stacking has energies comparable to AB-stacking and is at a local minimum. A similar result is obtained when the fully relaxed AB$^\prime$-stacking is used as the reference structure. Hence, even though AB-stacking corresponds to the global minimum, it is conceivable that the bilayer exfoliated at room-temperature is kinetically trapped in this AB$^\prime$-stacking order, on cooling. This is consistent with the lack of experimental reports of a structural transition on cooling.

We also calculate the interlayer exchange energy defined as the difference in energy between the FM and layered-AFM spin configurations, for each shifted configuration in Fig.~\ref{fig:exc}(a). We find that while AB-stacking strongly prefers FM, AB$^\prime$-stacking prefers AFM, albeit, weak [see Fig.~\ref{fig:exc}(b)], and is at the phase boundary of FM and AFM~\footnote{The magnetic ground state of AB$^\prime$-stacking is sensitive to the choice of $U$. For instance, $U$ of 2~eV results in a FM ground state. This is expected given that AB$^\prime$-stacking is at the phase boundary of AFM and FM.}. The position of AB$^\prime$-stacking in the magnetic phase diagram closer to the FM-AFM boundary is consistent with both the observed giant ME effect~\cite{Huang18p544,Jiang18p41563}, and the ease with which the magnetism can be changed by electrostatic doping.
 
Further, Fig.~\ref{fig:exc}(b) shows a strong coupling between the stacking order and the magnetism. Apart from the bulk-like stacking orders (the AB-stacking and the AB$^\prime$-stacking orders), we identify other magnetically important stacking orders. While AA-stacking and AB-stacking prefer FM, we find that there are other regions which strongly favor AFM.  
Among those, there are two symmetry inequivalent stacking orders which have the same point group symmetries as the bilayer from the room temperature phase (AB$^\prime$-stacking). They are labelled as AC$^\prime$ and AB$^\prime_1$ in Fig.~\ref{fig:exc}, and corresponds to a fractional lateral shift of the bottom CrI$_3$ monolayer unit by [2/3,0] and [1/6, -1/6] with respect to the AB-stacking, respectively. Fig.~\ref{fig:exc}(a) suggests that both AC$^\prime$- and AB$^\prime_1$-stacking are energetically unfavorable. However, these strong AFM centers are relevant in a Moir{\'e} superlattice.

Are there other interesting stacking orders? To answer this we calculated the stacking energy and the interlayer exchange energy for the full 2D space of lateral shifts, by performing a rigid shift of one CrI$_3$ layer parallel to the basal plane of the second layer, starting with the fully relaxed AB-stacking. A similar result is obtained when the fully relaxed AB$^\prime$-stacking is used as the reference structure. We have chosen a $6\times6$ grid to perform the lateral shifts. Fig.~\ref{fig:exc}(c) and (d) shows the stacking energy landscape and the interlayer exchange energy landscape, respectively, for the full space of lateral rigid shifts. AB-stacking still remains the global minimum, with AB$^\prime$-stacking orders having a comparable stacking energy. AB$^\prime$-, AC$^\prime$- and AB$^\prime_1$-stacking orders come in groups of three, as dictated by symmetry [see Fig.~\ref{fig:exc} (d)]. The strong coupling between the stacking order and the magnetism is also evident. As all the above-identified structures are included in the line-cuts along the [100]- and [1$\bar{1}$0]-directions, we restrict the discussions to these high-symmetry directions, which are representative of the full lateral stacking configuration space.

\begin{figure}
\includegraphics[width=1.0\columnwidth]{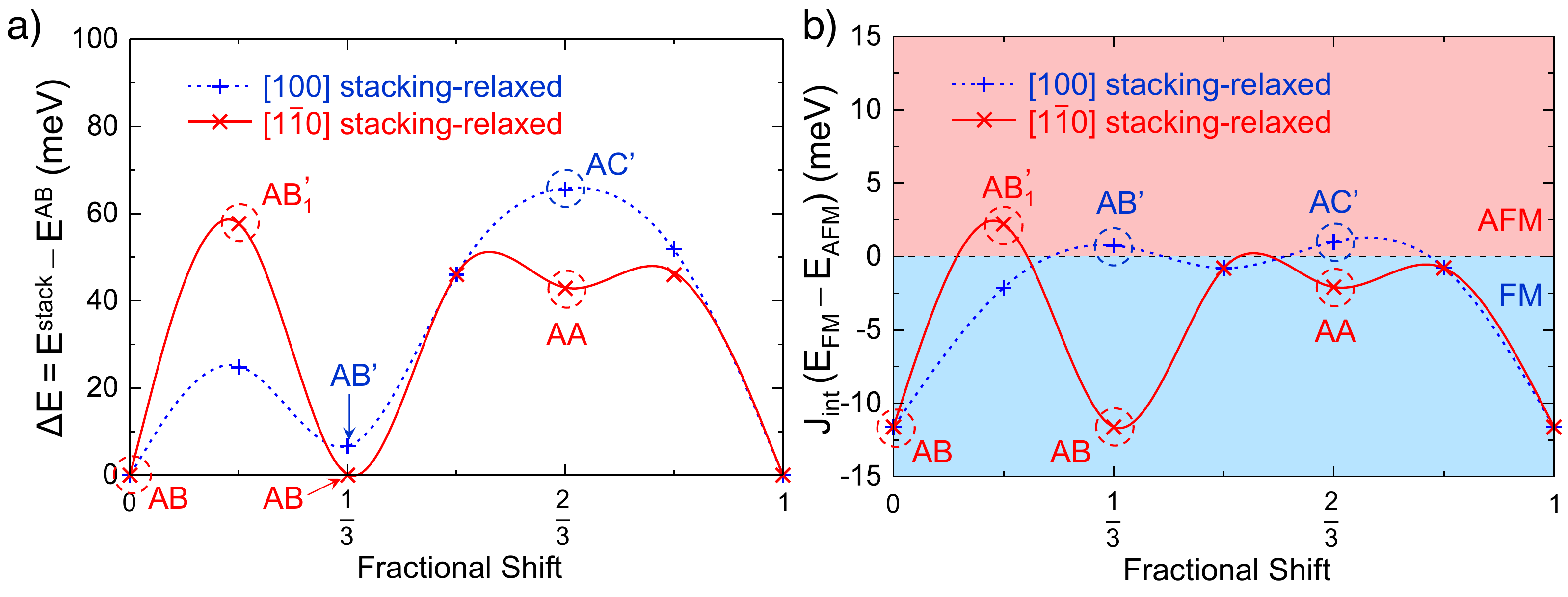}
\caption{\label{fig:int} Effect of stacking-constraint relaxation on (a) the stacking energy and (b) the interlayer exchange energy for the high-symmetry [100]- (dotted blue line) and [1$\bar{1}$0]-directions (solid red line). The identified high-symmetry stacking orders are labeled.}
\end{figure}

So far, we have only considered rigid shifts of one monolayer with respect to the other, using fully relaxed AB- and AB$^\prime$-stacking orders as the reference. Exfoliated thin-films transferred onto a substrate can further relax, retaining the strain and stacking order~\cite {Lee17p887,Jung17p085442}. Hence, we performed stacking-constraint relaxation along the high-symmetry directions. Fig.~\ref{fig:int} shows the effect of this on (a) the stacking energy and (b) the interlayer exchange energy for the high-symmetry [100]- (blue dotted line) and [1$\bar{1}$0]-directions (red solid line). The main effect of the relaxation is to reduce the energies of the high stacking energy configurations, without qualitatively affecting the overall trends [see Fig.~\ref{fig:int}(a)]. The is accompanied by a change in the interlayer distances (see AC$^\prime$ and AB$^\prime_1$ in Table~\ref{tab:tablecry}). The relaxed interlayer distances for AB-stacking and  AB$^\prime$-stacking compares well with the bulk experimental interlayer distances~\cite{McGuire15p612}. While AB$^\prime$-stacking remains a local minimum after the relaxation, the energy barriers surrounding it is reduced. 

The corresponding effect of the relaxation on the interlayer exchange energies is shown in Fig.~\ref{fig:int}(b). Similar to the stacking energies, the interlayer exchange energies are qualitatively unaffected. AB- and AA-stacking prefers FM with AB$^\prime$-, AC$^\prime$- and AB$^\prime_1$-stacking preferring AFM, just like in Fig.~\ref{fig:exc}(b).  
We thus conclude that the stacking order defines the interlayer exchange coupling for the above-identified structures. This strong spin-stacking coupling in bilayer CrI$_3$ provides a new route to tune the magnetic ground state between AFM and FM.

Next, we discuss the microscopic mechanism of the magnetic coupling. Because  the Cr atoms in CrI$_3$ are in a $d^3$ electronic configuration, and the top of the valence band is made up of a combination of the three $t_{2g}$ orbitals, the simplest model to describe the magnetism in bilayer CrI$_3$ includes only the nearest-neighbor intralayer and interlayer exchange interactions~\cite{sup-Sta-dep}. We find that the intralayer exchange is strongly FM (-2.2 meV/$\mu_B^2$) and dominates the interlayer exchange which is stacking dependent (0.04 meV/$\mu_B^2$ for AB$^\prime$-stacking)~\cite{sup-Sta-dep}. Therefore, CrI$_3$ bilayers can be viewed as two macroscopic spins with AFM interactions between them. The energy required to flip a spin involves not only overcoming the interlayer exchange energy but also the intralayer exchange energy as well. This suggests that the N\'eel temperature for the bilayer should be comparable to the Curie temperature of the monolayer, as reported by experiments~\cite{Huang17p270}.

\begin{figure}
\includegraphics[width=1.0\columnwidth]{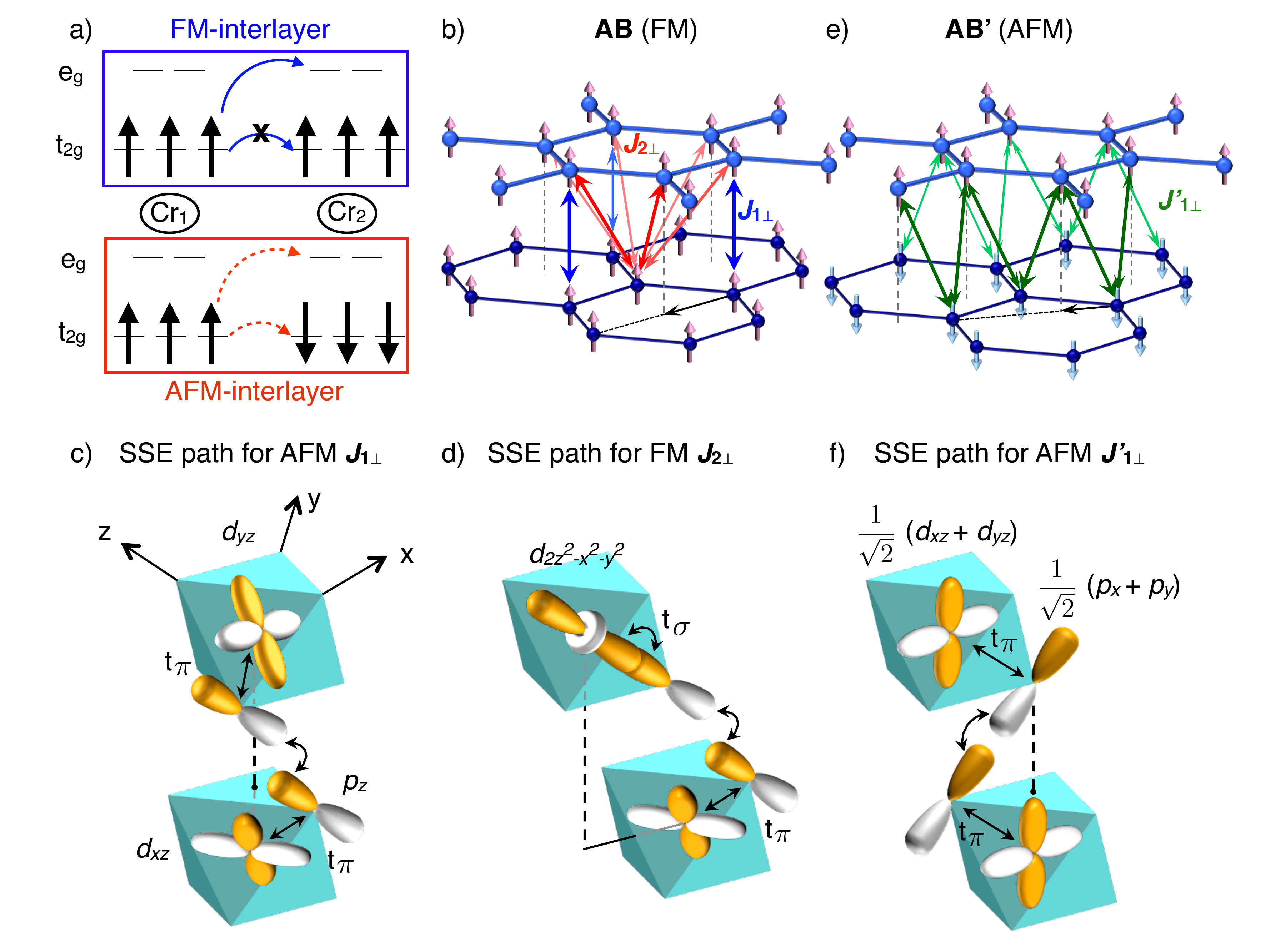}
\caption{\label{fig:se} The interlayer exchange. (a) A schematic of the orbital dependent interlayer SSE interactions. Hopping of the form $t_{2g}$-$t_{2g}$ is prohibited in FM exchange (blue), whereas this is allowed in AFM exchange (red). The interlayer Cr  nearest-neighbor ($J_{1 \perp}$, in blue) and the second-neighbor ($J_{2 \perp}$, in red) in (b) AB-stacking, and (e) the nearest-neighbor ($J^\prime_{1\perp}$, in green) in AB$^\prime$-stacking. Schematics for (c) the AFM SSE $J_{1 \perp}$ involving half-filled $t_{2g}$ orbitals, (d) the FM SSE $J_{2 \perp}$ between half-filled $t_{2g}$ orbital and empty $e_{g}$ orbital, and (f) the AFM SSE $J'_{1 \perp}$ between half-filled $t_{2g}$ orbitals.}
\end{figure}

As we are interested in the stacking-dependent magnetism, we focus on the interlayer exchange interaction. Fig.~\ref{fig:se}(a) shows a schematic of the different exchange interactions between Cr atoms in different layers. Hopping of the form $t_{2g}$-$t_{2g}$ is prohibited for FM alignment (blue), whereas this is allowed for AFM alignment (red). Therefore, $t_{2g}$-$t_{2g}$ hybridization leads to AFM.  On the other hand, hopping of the form $t_{2g}$-$e_g$ leads to an exchange coupling that is predominantly FM because of the local Hund coupling~\cite{Goodenough55p564,Goodenough58p287,Kanamori59p87}. All these interlayer Cr-Cr exchange interactions are mediated by the hybridization between the I $p$ orbitals, and therefore, SSE in nature~\cite{Ehrenberg98p152}. The stacking-dependent magnetism originates from a competition between the different interlayer orbital hybridizations.

Such a competition is most evident in AB-stacking. The interlayer Cr-Cr nearest-neighbors $J_{1 \perp}$ (blue) and the second-neighbors $J_{2 \perp}$ (red) for AB-stacking is shown in Fig.~\ref{fig:se}(b). $J_{1 \perp}$ is dominated by virtual excitations between Cr half-filled $t_{2g}$ orbitals, as shown schematically in Fig.~\ref{fig:se}(c), and induces an AFM coupling~\cite{Anderson50p350}. In stark contrast, $J_{2 \perp}$ is dominated by a virtual excitation between the Cr half-filled $t_{2g}$ orbitals and the empty $e_g$ orbitals resulting in a net FM coupling, as shown in Fig.~\ref{fig:se}(d). AB-stacking has one $J_{1 \perp}$ bond and sixteen $J_{2 \perp}$ bonds per unit cell, as listed in Table~\ref{tab:tablecry}. Hence, the second-neighbor FM interlayer SSE dominates the AFM nearest-neighbor interlayer exchange, making AB-stacking strongly FM~\footnote{We calculated the hopping amplitude for $t_{2g}$-$t_{2g}$ and the $t_{2g}$-$e_g$ using first-principles based  Wannier90 to explicitly check this~\cite{PhysRevB.56.12847,Souza01p035109,Mostofi08p685}.}. 

\begin{table}[b]
\caption{\label{tab:tablecry} The interlayer nearest-neighboring (NN) and second-neighboring distances (2nd-NN) between Cr atoms, with the corresponding coordination number per unit cell in square brackets for the different stacking orders from a rigid shift of the AB-stacking (stacking-constraint relaxed). The shortest interlayer I-I distance is also listed.}
\begin{tabular}{|c|c|c|c|c} \hline
Stacking & Cr-Cr NN (~\AA~) & Cr-Cr 2nd-NN (~\AA~) &  I-I (~\AA~)\\
\hline
AB & 6.7 (6.7) [1] & 7.8 (7.8)~[16] & 4.2 (4.2) \\
AB$^\prime$ & 7.1 (7.1)~[6] & 8.1 (8.2)~[6] & 4.2 (4.2) \\
AC$^\prime$ & 7.1 (7.7)~[6] & 8.1 (8.7)~[6] & 3.6 (4.3) \\
AB$^\prime_1$ & 7.0 (7.5)~[2] & 7.5 (8.0)~[8] & 3.7 (4.2) \\
AA & 6.7 (7.0)~[2] & 7.8 (8.0)~[10] & 4.1 (4.3)\\
\hline 
\end{tabular}
\end{table}

For each stacking order, the ground state magnetism is determined by the competition between the different orbital-dependent AFM and FM exchange interactions. A lateral shift of one monolayer with respect to the other breaks the interlayer hybridization between I $p$ states and generates new ones. This difference is evident while comparing AB$^\prime$-stacking [see Fig.~\ref{fig:se}(e)] to the AB-stacking. For the AB$^\prime$-stacking, while the nearest-neighbor coordination number is increased compared to the AB-stacking the second-neighbors coordination number is decreased [see Table 1]. The combined effect is to reduce the strength of the FM exchange interactions while increasing the AFM exchange, resulting in an AFM ground state for AB$^\prime$-stacking. Fig.~\ref{fig:se}(f) shows a schematic of the dominant AFM SSE interaction for AB$^\prime$-stacking involving the $t_{2g}$ orbitals.

This stacking-dependent magnetism mediated by the I octahedrons is not only applicable to bilayer CrI$_3$ but also for other octahedrally coordinated 2D-magnets, including Cr$X_3$ ($X =$ Cl, Br, I)~\cite{Handy52p891,Bengel95p95,Sivadas15p235425,McGuire17p064001}. Further, our results have important implications in the Moir\'{e} physics of 2D-materials~\cite{Dean13p598,Jeil14p205414,Cao2018p43}. A twisted stacking of bilayers immediately results in regions with spatial variations in the local stacking order. For bilayer CrI$_3$, the local stacking order defines the length scale of the magnetic Moir\'{e} patters~\cite{Jeil14p205414}. While both AC$^\prime$- and AB$^\prime_1$-stacking were previously identified as energetically unfavorable [see Fig.~\ref{fig:exc}], they form the centers for AFM in the twisted geometry. Similarly, the AB-stacked regions will form the FM centers. Therefore, our results on stacking-dependent magnetism have a direct impact on the Moir\'{e} physics of two-dimensional magnets.
 
\begin{table}
\caption{\label{tab:table1} The form of the symmetry allowed linear ME tensor for bilayer CrI$_3$ for different stacking orders and magnetic orders. The corresponding point group is also labeled in brackets. A linear ME effect is prohibited for all the FM configurations, irrespective of the number of layers and stacking order, as long as inversion is a symmetry of the magnetic space group.
}
\begin{tabular}{|c|c|} \hline
Magnetic and stacking order & Linear ME coefficient \\
\hline
AFM: AB-stacking (S$_6$)& $\begin{pmatrix}
    \alpha_{xx}       & \alpha_{xy} & 0 \\
   -\alpha_{xy}       & \alpha_{xx} & 0 \\
    0       & 0 &  \alpha_{zz}
\end{pmatrix}$
\\
AFM: AB$^\prime$-stacking (C$_{2h}$)& $\begin{pmatrix}
    \alpha_{xx}       & 0 &\alpha_{xz} \\
    0       & \alpha_{yy} & 0 \\
    \alpha_{zx}       & 0 &  \alpha_{zz}
\end{pmatrix}$
    \\
AFM: AA-stacking (D$_{3d}$)& $\begin{pmatrix}
    \alpha_{xx}       & 0 & 0 \\
    0       & \alpha_{yy} & 0 \\
    0       & 0 &  \alpha_{zz}
    \end{pmatrix}$
    \\
 AFM: General stacking (C$_i$) & $\begin{pmatrix}
    \alpha_{xx}       &  \alpha_{xy} &\alpha_{xz} \\
    \alpha_{yx}      & \alpha_{yy} & \alpha_{yz} \\
    \alpha_{zx}       & \alpha_{zy} &  \alpha_{zz}
\end{pmatrix}$
\\ \hline
\end{tabular}
\end{table}

Finally, to experimentally verify the predicted stacking-dependent magnetism, and to characterize the magnetic ground state of bilayer CrI$_3$, we propose the following strategy. Since both time-reversal and inversion symmetry are broken in all AFM configurations, a linear magnetoelectric (ME) effect is allowed by symmetry. But, the specific form of the linear ME tensor $\alpha_{ij} = \partial M_i/\partial E_j$ is determined by the crystalline symmetries itself, and therefore, is strongly dependent on the stacking order.  Table~\ref{tab:table1} lists the symmetry allowed form for the linear magnetoelectric (ME) effect for these different AFM stacking orders. For instance, the form of the linear ME tensor for AB-stacking and AB$^\prime$-stacking are very different. While the former allows a toroidal linear ME ($\alpha_{xy}$) in addition to a transverse linear ME ($\alpha_{xx}$) and longitudinal linear ME ($\alpha_{zz}$), the anti-symmetric toroidal term is absent in the latter. Instead, an off-diagonal term, $\alpha_{xz}$, which is independent of $\alpha_{zx}$, is symmetry allowed. Thus, by doing a systematic measurement of the various components of magnetization produced in response to a gate-voltage, one can distinguish between the various stacking order in bilayer CrI$_3$.

In summary, we have shown that magnetism in CrI$_3$ bilayers is strongly stacking-dependent, and could be changed between FM and AFM by changing the stacking order.  This is a general phenomenon that should exist for a broad class of 2D honeycomb magnets. The sensitive dependence of the magnetic ground state on the local bonding environment suggests great potential for mechanically tuning the magnetism.

NS and CJF are supported by NSF through the Platform for the Accelerated Realization, Analysis, and Discovery of Interface Materials (PARADIM) (DMR-1539918), SO is supported by the U.S. Department of Energy, Office of Science, Basic Energy Sciences, Materials Sciences and Engineering Division, DX and XX acknowledge the support from DOE BES Pro-QM EFRC (DE-SC0019443).
The authors thank Kin Fai Mak, Jie Shan, David Muller and Darrell Schlom for interesting discussions and insightful comments. NS thanks Valentino R. Cooper for sharing the input files for bulk CrI$_3$. NS would also like to thank Gerhard H. Olsen, Bet\"{u}l Pamuk, Matthew W. Daniels, Hena Das and Jisha V.M for various discussions and help with visualization.  NS thanks Cornell University Center for Advanced Computing for the computing time.

\textit{Note added}.---Recently, two related papers~\cite{Jiang18p01,Soriano18p00} have studied the stacking-dependent magnetism in CrI$_3$ bilayers, but only focused on the AB and AB$^\prime$-stacking. The stacking-dependent magnetism has also been discussed in the Supplementary Note of Ref.~\citenum{Wang18p2516}.

Supporting Information Available: See Supporting Information for additional structural and magnetism information.

\providecommand{\latin}[1]{#1}
\providecommand*\mcitethebibliography{\thebibliography}
\csname @ifundefined\endcsname{endmcitethebibliography}
  {\let\endmcitethebibliography\endthebibliography}{}

\newpage
\begin{figure}
\includegraphics[width=1.0\columnwidth]{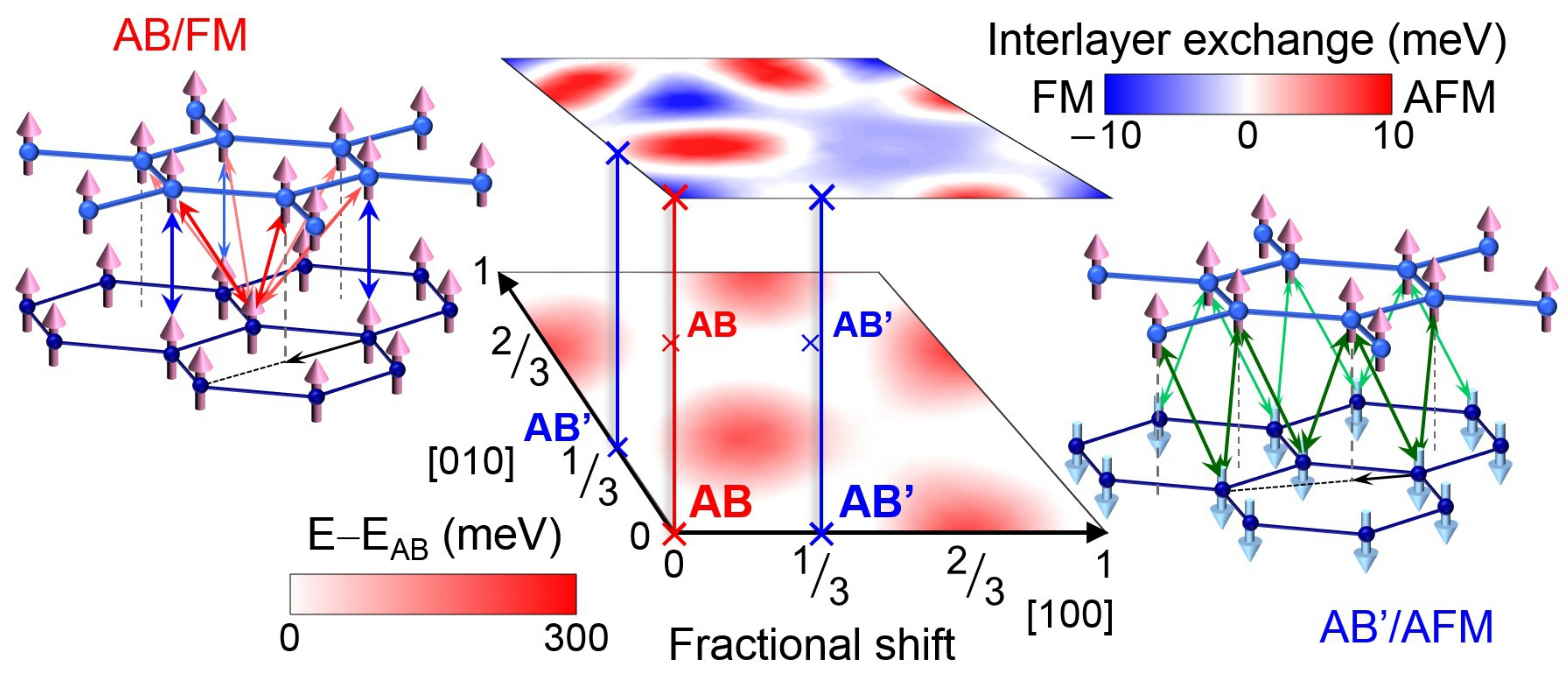}
%\caption{The stacking energy and the interlayer exchange energy as a function of lateral shift, with respect to the AB-stacking. By changing the interlayer stacking order one can tune the interlayer exchange interaction between antiferromagnetic and ferromagnetic.}
\end{figure}

\end{document}